\documentclass[runningheads,a4paper]{llncs}

\usepackage{amssymb}
\usepackage{graphicx}
\usepackage{url}

\usepackage[llncs,crop]{llncsconf}
\conference{}
\llncs{This is a post-peer-review, pre-copyedit version of an article published in Lecture Notes in Computer Science (LNCS, volume 9923).}{1}
\llncsdoi{10.1007/978-3-319-45480-1_26}

\urldef{\mailsa}\path|{marc.zeller, daniel.ratiu, kai.hoefig}@siemens.com|

\begin{document}

\mainmatter  

\title{Towards the Adoption of Model-based Engineering for the Development of Safety-critical Systems in Industrial Practice}

\titlerunning{Towards the Adoption of Model-based Engineering}

%
%
\author{Marc Zeller\and Daniel Ratiu\and Kai H{\"o}fig}
%

\institute{Siemens AG, Corporate Technology\\
Otto-Hahn-Ring 6, 81739 Munich, Germany\\
\mailsa}

%
%

\maketitle

\begin{abstract}
Model-based engineering promises to boost productivity and quality of complex systems development. In the context of safety-critical systems, a traditionally highly regulated and conservative domain, the use of models gained importance in the recent years.
In this paper, we present a set of practical challenges in developing safety-critical systems with the help of several examples of development projects that belong to different application domains.
Following this, we show how could the adoption of model-based engineering for the development of safety-critical systems cope with these challenges.
\end{abstract}

\section{Introduction}
\label{sec:introduction}
Model-based development is currently one of the key approaches to cope with increasing development complexity in general. 
Particularly the development of today’s safety-critical systems underlies a series of legislative and normative regulations making safety the most important non-functional property of embedded systems. 
Applying model-based approaches during the development of complex products means the use of adequate models for different aspects of the system. Such models ease the development, increase the quality and enable a systematic reuse. This has the potential to help the industry to meet even tighter deadlines for new products and decrease the costs. 

Along with the growing system complexity the effort needed for safety assessment is increasing drastically in order to guarantee the high quality demands. However, this trend is contrary to industry's aim to reduce development costs and time-to-market of new products.
The use of models would help along two directions.
Firstly, it makes safety engineering as a standalone sub-task of system development more efficient. Secondly, and even more important, this is an essential step towards a holistic model-based development approach which closes the gap between functional development and safety assessment. Reusing development models for safety analyses and feeding back the results of safety analyses in the development models is a key step for reaching synergies.

Although a large body of research results about using model-based development for safety-critical system already exists, they did not found their way into the industrial practice yet. In this paper, we outline the current practice in developing safety-critical systems and derive a set of challenges. Based on this, we describe how would the adoption of Model-Based Engineering (MBE) in the development of safety-critical systems cope with these challenges. Therefore, we present how can models help to assess that a system is sufficiently safe (models for safety) and how models can be applied for the development of safety-critical systems (safety for models).

\section{Safety Assessment at a Glance}
\label{sec:safety_assessment}
\begin{figure*}[t]
	\centering
	\includegraphics[width=0.7\linewidth]{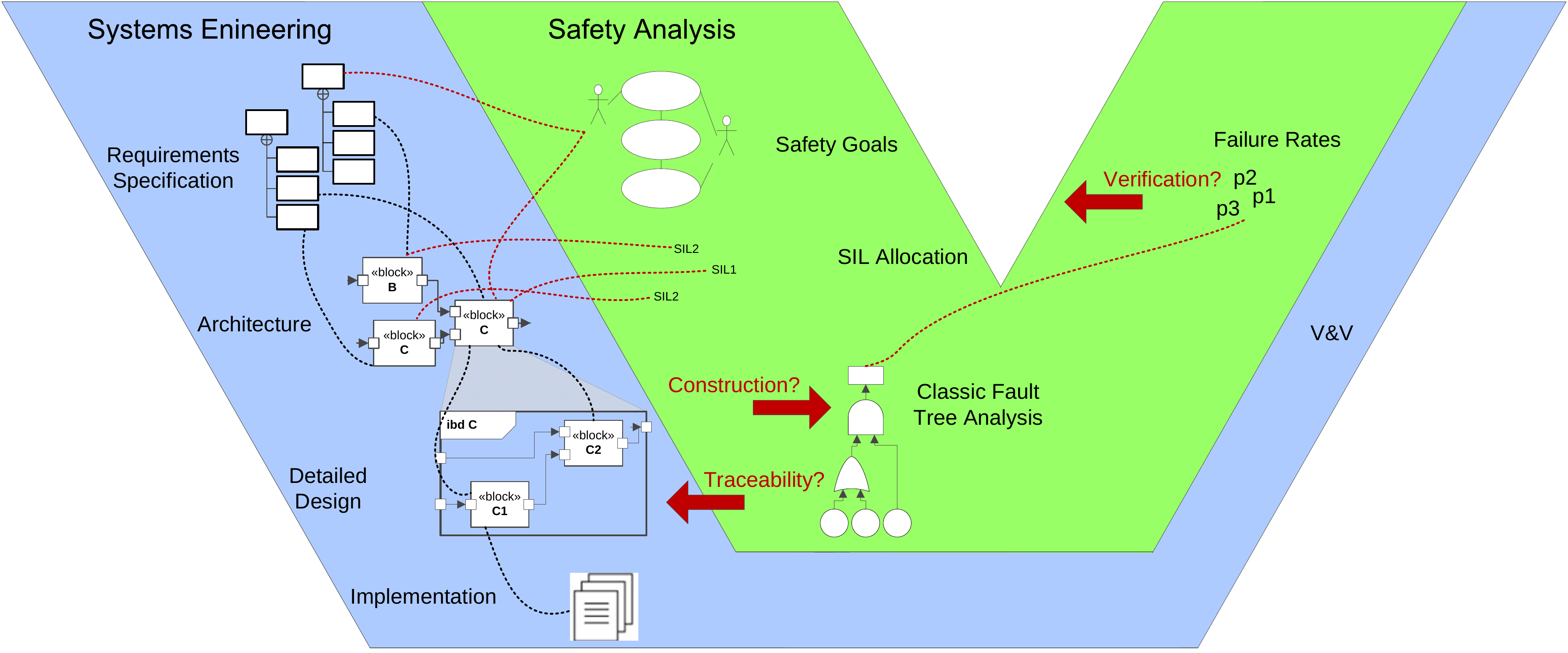}
	\caption{Development of safety-critical systems in current industrial practice}
	\label{fig:safety_overview}
\end{figure*}

The goal of the safety assessment process is to identify all failures that cause hazardous situations and to demonstrate that their probabilities are sufficiently low. In the application domains of safety-relevant systems, the safety assurance process is defined by the means of safety standards (e.g.~IEC 61508 \cite{ElectrotechnicalCommission1998}).
Although each domain has its own standards and regulation, the safety assessment includes a generic set of activities which are related to the system engineering process (see Fig.~\ref{fig:safety_overview}).

As a first step, safety goals are defined according the system requirements. Based on the system requirements the architecture of the system is designed. After performing a hazard and risk analysis, \emph{Safety Integrity Level (SIL)} information is obtained and allocated to the elements in the system architecture (e.g.~the components of the systems). In the next step, the detailed design of the system is built which is the basis for the implementation of the system. Based on the detailed system design, the safety engineer is developing a safety analysis for the specified system.
Traditionally, safety analysis consists of bottom-up safety analysis approaches, such as \textit{Failure Mode and Effect Analysis (FMEA)}, and top-down ones, such as \textit{Fault Tree Analysis (FTA)}, to identify failure modes, their causes, and effects with impact on the system safety.
The result of a (quantitative) analysis is a set of failure rates for the hazardous events which are used for the verification of the safety requirements defined in the safety case.

\section{Current Practice in Developing Safety-critical Systems}
\label{sec:current_practice}
In the following, we present four examples which illustrate the current practice of developing safety-critical systems in different domains at Siemens.

\subsection{Example 1: Modular Certification for Trains}
\label{sec:train}
Traditionally, trains have high requirements in terms of safety.
Modern trains are built using a modular platform concept. Such a platform concept enables the manufacturer to build various train configurations in a flexible way and reuse certain parts. 

Currently, a \emph{clone and own} strategy is used in order to build a new trainset based on the existing platform. However, every new trainset developed based on the same platform requires an individual certification. Even more, the certification needs to be granted by different certification authorities of the countries in which the trains will run.
Furthermore, in the railway domain, there is a substantial legacy and constraints imposed by the existing infrastructure. For instance, there are different train protection systems in use by railways across Europe. These control systems have both on-board components and side-track components which must interoperate flawlessly in order to ensure safe operation.

In current practice, model-based development is applied to isolated sub-systems of trains. The system development and safety assessment are mainly based on multiple specification and analysis documents.
However, the modularization of specific parts of a train enables their reuse for multiple trainsets. But changing one specific part of the train impacts multiple analysis documents. Thus, the adjustment of the safety analysis is a very time-consuming and complex manual task.

\subsection{Example 2: Reusable System for Industry Automation}
\label{sec:industrial_automation}
In the industry automation domain, compositional system development fosters individual solutions for customers with high potential for modular certification.
Thereby, every specific system consists of an individual set of solvers, sensors as well as actuators of different types and vendors. Each system is an individual combination of parts according to the customer's requirements. The architecture of industry automation systems is very flexible in terms of involved sensors and actuators. Moreover, the system is composed by reusing standardized components from a repository.

In current projects use, safety assessment is based on reusable certificates for the quality of the process. Process-based certification involves organization aspects, qualification of involved personnel and proof of quality goals for system elements.
However, with the growing system complexity process-based certification is becoming more and more expensive. Moreover, it does not support the compositional and flexible way current systems in industry automation are built.

To enable efficient product-based certification of the individual system, modular safety assessment as well as the systematic reuse of safety artifacts must be possible. Moreover, the safety assessment processes of the industry automation system should be embeddable into customers' certification processes for the overall manufacturing system.

Model-based development is solely applied for describing the components functionality of PLCs by using domain specific modeling languages (e.g.~Continuous Flow Charts, CFCs).

\subsection{Example 3: Medical Devices}
\label{sec:medical_devices}
Modern medical devices realize complex safety-critical functionalities due to multiple system states, placing of the machinery within the clinical environment and high sensitivity requirements. Requirements for healthcare systems in terms of safety are single-fault protection and partial fail-operational by two options: First, risk avoidance by bringing the system in a safe state. Second, providing independent redundancy.

In the current industrial practice, manually maintained tables are used to calculate failure rates, separate failure classes and to guide the safety analysis process and show that a medical system is sufficiently safe. The reuse of individual component or sub-systems is managed manually by copy-and-paste and cell references.

However, with the growing complexity of the functions of medical devices, which involve even larger circuits with increased reuse, more sophisticated methods and tools are needed for safety assessment in order to fulfill high quality demands in this domain and meet fail-operational requirements in the future. 

\subsection{Example 4: Future Automotive ICT Platform}
\label{sec:race}
Today's vehicles are filled with more information and communication technology (ICT) than ever before. A paradigm shift from the array of control units used today to a flexible set of software-implemented features stored on just a few central platform computers enables a cost-effective way to implement current as well as novel functionalities. Such an architecture, e.g.~developed in the German national funded project RACE\footnote{\url{http://www.projekt-race.de/en/}}, provides a central platform computer concept with fail-operational behavior.
Moreover, the platform aims to offer plug-and-play capabilities to easily enhance or integrate new features and components while the car is in the field.
Therefore, run-time (re-)qualification of the system w.r.t.~safety is the central future business case to ensure that the impact of the extension has no negative results.

In current practice, a fault containment region-based analysis provides reus\-able hardware failure rates for a later qualification of specific functionalities planed to run on the central ICT platform w.r.t.~safety \cite{hiefig2014a}.

However, this approach for system qualification is solely used during the development to assess the system in terms of safety. Run-time (re-)qualification based on the current approach is not yet possible.

\section{Challenges}
\label{sec:challenges}
The current practice in developing safety-critical systems in the context of industrial projects faces the following challenges w.r.t.~safety assessment:

\subsection{Efficient Construction of Safety Analyses}
In current practice, constructing a safety assessment (e.g.~using fault trees or FMEAs) and maintaining its quality through the development is a challenging and time-intensive manual task (see Sec.~\ref{sec:current_practice}). With the increasing system complexity the manual construction of a safety analysis model for an entire system is becoming very hard or even unfeasible.
Moreover, incremental and iterative development processes used in industry require safety analyses to be performed along the complete design process and provide immediate feedback to the system engineers about the safety aspects of the systems being developed.

To perform safety analysis efficiently in large-scale industrial projects, methods are required to construct safety analysis models in a structured way based on the information available in the detailed system design.

\subsection{Evolution of the System Design}
During most industrial development projects, change requests can come from various stakeholders such as the client, certification authorities or development teams of the different sub-projects. But changes can also be a part of a development strategy, if an existing product can be evolved in a new system in an incremental manner with small changes and adjustments (e.g.~in the development of trains based on a platform concept, see Sec.~\ref{sec:train}).
In case of modifications of the system design during the development process, the safety analysis must be adapted accordingly to guarantee that the results of the safety analysis are still valid. Since traceability between the artifacts in the system design and the safety assessment is solely achieved manually in current practice, each change within the system design results in time-consuming manual adjustment performed by the safety engineer.
For instance, after each modification all FMEA tables or fault trees of the system must be completely reviewed and all parts affected by the modification must be adapted.

In order to decrease the time-consuming adaptation of the safety analyses, traceability between the elements in the safety analysis and the related elements in the system design must be established \cite{schultz2011model}. Moreover, automated synchronization of the safety analysis model with changing system design in a continuous manner is needed.

\subsection{Systematic Reuse of Safety Artifacts}
In industrial practice, developers often have existing development artifacts which are reused to compose a system (see Sec.~\ref{sec:train} or Sec.~\ref{sec:industrial_automation}). Such development artifacts are for instance stored within a repository and put together as a new configuration for a product.
This compositional development strategy allows automated system construction from preexisting building blocks. With the motivation for reusing the results of safety analyses of existing development artifacts, the safety assessment needs to be aligned with compositional system design. Hence, safety artifacts on the granularity of component-level must be exchangeable within the safety analysis model.

To enable systematic reuse the safety artifacts related to system components should be stored within a repository and used to compose a system-wide safety analysis model. The composition of the safety artifacts should be performed at best in an automated way.

\subsection{Seamless Traceability}
In practice, the results of the safety analysis process cannot be mapped easily with the systems' safety goals, since their relation is often not clearly traceable and maybe ambiguous. Therefore, the verification of the systems' safety goals with the results of the safety analyses is a complex task itself.
This is a challenge in the development of safety-critical systems in large-scale industrial projects across all domains within Siemens (see Sec.~\ref{sec:current_practice}).
In order to enable an unambiguous mapping between the safety goals and the safety analyses results, seamless traceability between the safety goals, system specification, and safety analyses must be established.

For an efficient connection of different artifacts (e.g. specification, high-level design, and low-level design) within the system engineering process, information should be integrated automatically without the manual establishment of traceability links. 

\subsection{Automated (Re-)Qualification}
Adaptations and modifications of an embedded system are traditionally performed solely during the development. However, there is a strong trend to build open and adaptive system platforms (see the example of a future automotive ICT platform in Sec.~\ref{sec:race}).
These systems can be enhanced during run-time with novel functionalities and may be coupled temporary with other systems which dissolve and give place to other configurations.
The key problem in assessing the safety of such systems is that the configurations over its lifetime are unknown and potentially infinite. State-of-practice safety analysis techniques are currently applied during development and require an a priori knowledge of the configurations that provide the basis of the analysis of system. Such techniques are not applicable to open, adaptive systems that build up a new configuration at run-time.
Therefore, safety analyses must be applicable to assess novel system configurations ad-hoc during run-time in an automated way. Such that the adaptation or modification of the system in the field can be assessed in terms of safety.

\section{Leveraging Models in the Development of Safety-critical Systems}
\label{sec:mbse}
In this section, we show how to cope with the previous mentioned challenges by using models for the assessment that a system is sufficiently safe (Sec.~\ref{sec:modelbasedsafety}) as well as using models in the development of safety-critical systems (Sec.~\ref{sec:safetyformodels}). 

\subsection{Models for Safety}
\label{sec:modelbasedsafety}
The idea of model based safety assessment is to support automatic generation of classical safety artifacts such as fault trees or FMEA tables from system models \cite{Rauzy2002,papadopoulos04,Majdara2009}. Therefore, the system models are often annotated with failure propagation models to construct the safety artifacts.
Examples for such an approach are HiP-HOPS (Hierarchically Performed Hazard Origin and Propagation Studies) \cite{Papadopoulos1999} or Component Fault Trees (CFTs) \cite{Kaiser2003}.
These failure propagation models are commonly combinatorial in nature thus producing static fault trees. This is also driven by the industrial need to certify their system with static fault trees or FMEA tables. 
Only rarely more advanced safety evaluation models, such as Dynamic Fault Trees (DFTs) \cite{159800}, Generalized Stochastic Petri Nets (GSPNs) \cite{AjmoneMarsan:1984:CGS:190.191}, State-Event Fault Trees (SEFTs) \cite{Kaiser2004},
or Markov models \cite{IEC1995-2003}, exist in practice.
Besides annotating the architecture specification, there are also approaches in current research to synthesize safety artifacts via model checking techniques (e.g.~FSAP/NuSMV-SA \cite{Bozzano2003b}). However, such approaches have not found the way yet in the current industrial practice.

\subsection{Safety for Models}
\label{sec:safetyformodels}
Model-based development aims to address the high complexity of current systems by the use of adequate and rich models through all development phases from requirements engineering, to design, implementation, integration and deployment. Models are envisioned to be used at different granularity levels: Abstract models describe the entire system, and subsequently more concrete models are used at sub-system level until finest granular models are used at the component level. Thereby, the high-level models are kept in sync with lower level models. In an ideal world, the entire development is supported by a seamless and deeply integrated set of adequate models that address development concerns \cite{5420030}.

However, the adoption of model-based development in practice varies strongly between different industrial application domains.
In general, the current adoption and benefits of using model-based development is by far not reaching the promises given by the research community (see also \cite{2009_fieber_assessing_usability_of_mdd_in_industrial_projects}):

\emph{Requirements} are very weakly modeled if at all -- they are written using plain natural language text or they are captured in a hierarchical tree-like structure like that provided by DOORS. Besides the natural language text, requirements should be associated to meta-data such as the "safety integrity level" (for safety requirements). 

\emph{Architectures} are described informally with the help of "boxes and lines" pictures drawn in tools like MS Visio. When the architecture is modeled, it is done with SysML tools, like MagicDraw or Enterprise Architect. Many times the architecture is described in a hierarchical manner like systems consisting of subsystems, etc. The behavior specification of system-level abstract components is not a common practice (e.g. annotating interfaces of components with invariants, specifying pre/post conditions in a machine readable form). After the initial design, keeping the architecture consistent with the system implementation during the life-cycle and across product families requires a huge manual effort and long review cycles. 

\emph{Behavior of atomic components} is implemented using models or code, such as C/C++. When control- or state-based-like algorithms are being developed, then one of the mostly used tools is Simulink/Stateflow. For more hardware-close functionality however like device drivers or communication protocols, the plain code is used instead of models. Different domains already use domain specific modeling languages for describing the components functionality like SIBAS (railway), Simulink (controls) or PLC (industry automation) in large-scale projects.

However, besides the control algorithm modeling activities, which are current practices across different Siemens business units, the use of models for the system development is rather sporadic. Furthermore, most of the times ad-hoc tool-chains are used which comprise and extend commercially available tools with specific customizations.

\subsection{Integrating MBE with Safety-critical Systems Development}
\label{sec:adaption}
Using both models to analyze a system in terms of safety and models for the development of safety-critical systems, the practical challenges in industry, as previously described in Sec.~\ref{sec:challenges}, can be addressed as follows:
\subsubsection{Efficient Construction of Safety Analyses}
Models used in safety analysis, such as CFTs or HiP-HOPS, annotate the system models with failure propagation models. This enables the construction of the safety analysis model in a structured way.
Due to the use of models in the assessment of functional safety, advances of MBE such as providing traceability, tool support and consistency checks can be utilized.
Moreover, model-based safety engineering approaches allow the (semi-)automated generation of safety artifacts such as Fault Trees or FMEAs, if the system design is specified by using models (e.g.~\cite{Rauzy2002,papadopoulos04}).
Hence, the use of models for development improves the efficient construction of safety analysis models, since they reuse the information available in the system design and offer a sound methodology. As a result, safety analyses may be applied more frequently during the entire product development process.

From industrial adoption viewpoint, the construction of safety analysis models from the information in the system design models must be clearly traceable and understood by developers. Moreover, system and safety engineers must still be able to work with methodologies and models, with which they are familiar.

\begin{figure*}[tb]
	\centering
	\includegraphics[width=0.7\linewidth]{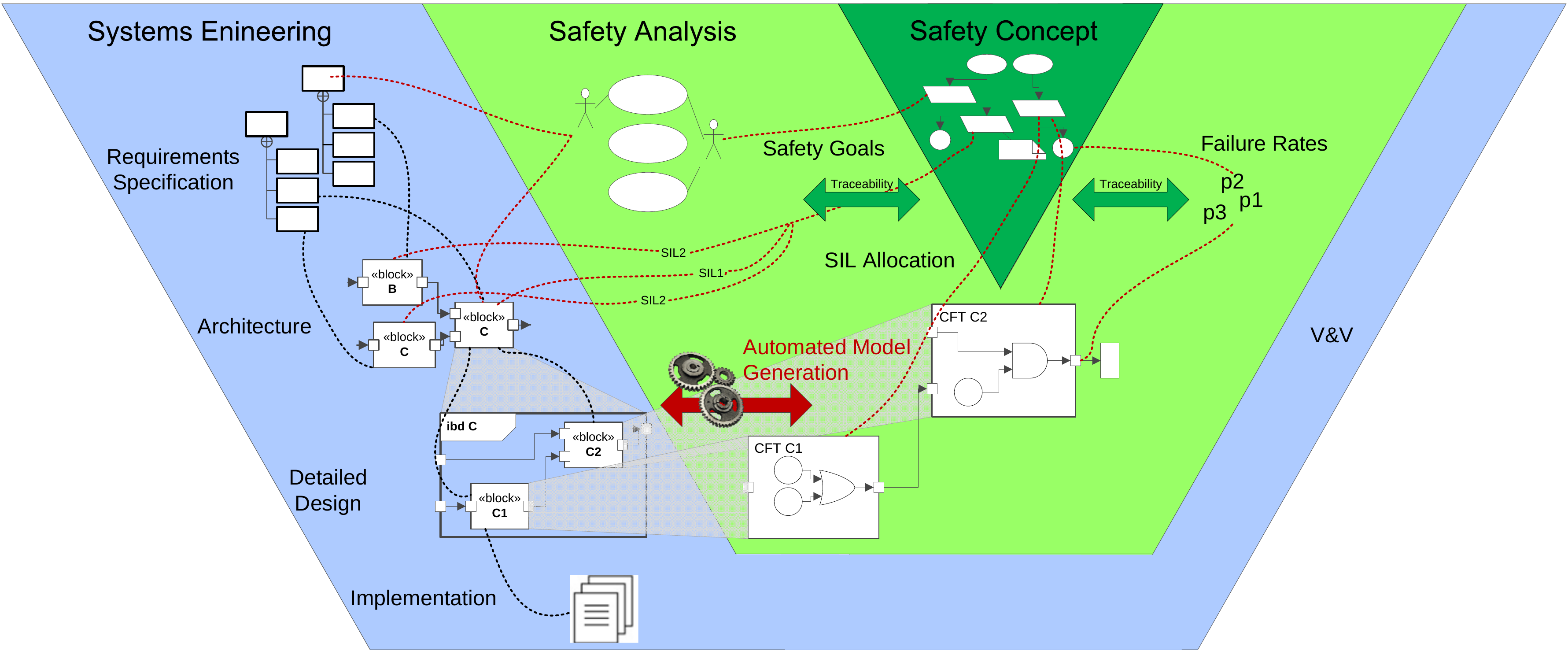}
	\caption{Adaption of MBE in the development of safety-critical systems}
	\label{fig:mbd_safety_overview}
\end{figure*}

\subsubsection{Evolution of the System Design}
When a model-based development strategy is used, traceability between system design and safety analysis artifacts is established.
Thus, the impact of changes in the system design is narrowed to encapsulated parts in the safety analysis model. The validity of the unchanged parts of the safety analysis is preserved.
For example, the modification of a specific system component does only affect the corresponding component within a CFT model.
Instead, in classic fault tree analysis, each fault tree must be reviewed manually whether to adjust certain sub-tress or not.

Also the information within both kinds of models must be consistent during the complete development process. For instance, if a certain system element is deleted or renamed, the safety analysis model must be adjusted accordingly. By using models in system design and safety assessment the synchronization can be performed (semi-)automatically to guarantee that the safety as well as the system engineers always work on consistent data.
Since the time-consuming maintenance of the safety analysis model is reduced significantly and safety analysis is kept in sync with the system design, safety analyses may be repeated during the complete development process. Thus, iterative development processes as used currently in industry can be supported in terms of safety assessment \cite{Zeller2016a}.

\subsubsection{Systematic Reuse of Safety Artifacts}
Since the models used in model-based safety assessment interlink safety with the system design artifacts, it is possible to reuse these safety artifacts in the safety assessment of different contexts. Hence, it is possible to construct a safety analysis model based on the reuse of preexisting parts and the specification of the newly created parts (compositional safety assessment).
In terms of top-down safety analysis, one possible direction is to use the CFT methodology and to establish a framework to synchronize with the system design model and to store and exchange specific CFT elements \cite{hoefig2015a}. Another direction is to enable reuse in bottom-up safety analysis, e.g.~by introducing model-based FMEA techniques \cite{hoefig2014}.
However, in order leverage compositional safety assessment in industrial practice, techniques for the automated composition of safety artifacts are need \cite{Moehrle2015}. First, as a preliminary approach, to give system engineers a first feedback w.r.t. system's safety in the early stages of the development. Second, to automate the system qualification in terms of functional safety.

\subsubsection{Seamless Traceability}
By the use of model-based approaches for safety analysis, traceability between system design elements (e.g.~components) and safety artifacts (e.g.~the failure propagation within a component) is established \cite{Kaiser2003,Papadopoulos1999}. Hence, it is possible to synchronize system design and safety artifacts and prevent inconsistencies during the development.
However, to be able provide an unambiguous relation of the results of the safety analyses and the systems' safety goals, we need to make the argumentation explicit by describing the safety goals and providing links to the analysis results which prove that the goal is fulfilled. Model-based approaches for constructing safety argumentation, such as the Goal Structuring Notation (GSN) \cite{kelly2004}, close this gap by providing links between safety goals, system design elements, safety analysis, and their results (see Fig.~\ref{fig:mbd_safety_overview}).
Thus, seamless traceability in the development of safety-critical systems is achieved by combining models for development, safety analysis and building safety concepts in a pragmatic way.
However, this is an intermediate step on the way towards the use of a \textit{holistic product model} which provides deep, coherent and comprehensive integration of requirements, specification, implementation, test/verification \& safety models \cite{5420030}.

\subsubsection{Automated (Re-)Qualification}
The safety of upcoming embedded systems cannot be fully assured prior to deployment (see Sec.~\ref{sec:race}). In order to assure the safety of such reconfigurable system, the degree of automation in safety assessment must be further increased.
Using models for the system design and the safety analysis provides a relation of system design elements and safety artifacts and enables the reuse of safety artifacts. By providing techniques to compose safety analysis automatically from preexisting building blocks, the (re-)qualification of the system in terms of safety can be automated.
Moreover, to enable the in-the-field safety assessment of a system, the methodology must be able to deal with system parts which provide no or incomplete information about its failure propagation. This is because upcoming embedded systems may interact spontaneously during operation including parts which are produced by different companies.
Therefore, methods are needed to automatically fill up empty safety analysis artifacts \cite{hoefig2015} in order to be able to perform a safety assessment of a system configuration, which is not know a priori.

\section{Related Work}
\label{sec:related_work}
%
Many papers (e.g.~\cite{Baker2005,Liebel2016}) discuss the challenges in MBE from an industrial practice. However, the specific characteristics in the development of safety-critical systems are not considered.

The use of models in safety assessment processes has gained increasing attention in research within the last decade \cite{McDermid2006,Lisagor2010,Lisagor2011,Sharvia2016}. But to our knowledge this is the first work which deals with the actual challenges that model-based safety engineering faces in industrial practice.

\section{Conclusion}
\label{sec:conclusion}
In this paper, we outline the current practice in developing safety-critical systems with the help of several examples of different business domains of Siemens. Based on this experience, we derive the challenges in the industrial practice.
Moreover, we describe how the adoption of MBE for the development of safety-critical systems can cope with these challenges from a practitioner’s viewpoint.
Therefore, we advocate that there is a dual perspective of the use of models in the context of safety-critical systems development. First, by using models for the assessment, that a system is sufficiently safe (models for safety). Second, using models for the design of safety-critical systems (safety for models). Only if these two perspectives are addressed jointly, models can leverage the development of safety-critical systems efficiently.



\bibliographystyle{IEEEtran}
\bibliography{references}
%

\end{document}